\documentclass[fp,twocolumn]{jpsj3}
\usepackage[dvipdfmx]{graphicx}
\usepackage{mathrsfs} 
\usepackage{txfonts}
\usepackage{bm}
\newcommand{\kf}{k_{\mathrm{F}}}
\newcommand{\ef}{\varepsilon_{\mathrm{F}}}
\newcommand{\as}{a_{s}}
\newcommand{\Tf}{T_{\mathrm{F}}}

\newcommand{\up}{\uparrow}
\newcommand{\dwn}{\downarrow}

\newcommand{\iom}{i\omega_n}
\newcommand{\inu}{i\nu_m}
\newcommand{\bp}{\bm{p}}
\newcommand{\bq}{\bm{q}}

\def\lesssim{\ \raise.3ex\hbox{$<$}\kern-0.8em\lower.7ex\hbox{$\sim$}\ }
\def\gesim{\ \raise.3ex\hbox{$>$}\kern-0.8em\lower.7ex\hbox{$\sim$}\ }

\title{Shear viscosity and Strong-Coupling Corrections in the BCS-BEC Crossover Regime of an Ultracold Fermi Gas}

\author{Daichi Kagamihara\thanks{dkagamih@rk.phys.keio.ac.jp}$^{1}$, Daisuke Inotani$^{2}$, and Yoji Ohashi$^{1}$}
\inst{$^{1}$Department of Physics, Keio University, 3-14-1 Hiyoshi, Kohoku-ku, Yokohama 223-8522, Japan\\
$^{2}$ Department of Physics, and Research and Education Center for Natural Sciences, Keio University, 4-1-1 Hiyoshi, Yokohama, Kanagawa 223-8521, Japan
}

\abst{We theoretically investigate the shear viscosity $\eta$ in the BCS-BEC crossover regime of an ultracold Fermi gas with a Feshbach resonance. Within the framework of the strong-coupling self-consistent $T$-matrix approximation, we examine how a strong pairing interaction associated with a Feshbach resonance affects this transport coefficient, in the normal state above the superfluid phase transition temperature $T_{\rm c}$. We show that, while $\eta$ diverges in both the weak-coupling BCS and strong-coupling BEC limits, it becomes small in the unitary regime. The minimum of $\eta$ is obtained, not at the unitarity, but slightly in the strong-coupling BEC side. This deviation is consistent with the recent experiment on a $^6$Li Fermi gas. In the weak-coupling BCS regime, we also find that $\eta$ exhibits anomalous temperature dependence near $T_{\rm c}$, which is deeply related to the pseudogap phenomenon originating form strong pairing fluctuations.}

\begin{document}
\maketitle
\par
\section{Introduction}
\par
Since the realization of the superfluid phase transition in $^{40}$K\cite{Regal:2004aa} and $^6$Li Fermi gases\cite{Zwierlein:2004aa,Kinast:2004aa,Bartenstein:2004aa}, thermodynamic properties of this system in the BCS (Bardeen-Cooper-Schrieffer)-BEC (Bose-Einstein condensation) crossover region\cite{Eagles:1969aa,Leggett:1980aa,Nozieres:1985aa,Sa-de-Melo:1993aa,Haussmann:1993aa,Haussmann:1994aa,Pistolesi:1994aa,Holland:2001aa,Timmermans:2001aa,Ohashi:2002aa} have extensively been studied\cite{Bloch:2008aa,Giorgini:2008aa,Chin:2010aa,Luo:2007aa,Luo:2009aa,Horikoshi:2010aa,Nascimbene:2010aa,Navon:2010aa,Nascimbene:2011aa,Sanner:2011aa,Sommer:2011aa,Meineke:2012aa,Ku:2012aa,Sagi:2012aa,Lee:2013aa,PhysRevX.7.041004}, by maximally using the advantage that a paring interaction in this system can be tuned by adjusting the threshold energy of a Feshbach resonance\cite{Chin:2010aa}. Furthermore, as a new research direction, non-equilibrium transport properties in the BCS-BEC crossover region have recently been discussed\cite{Cao:2011ab,Cao:2011aa,Sommer:2011aa,Brantut:2012aa,Brantut:2013aa,Elliott:2014aa,Elliott:2014ab,Bardon:2014aa,Joseph:2015aa,Trotzky:2015aa,Krinner:2016aa,Krinner_2017}. 
\par
Regarding this new direction, the shear viscosity $\eta$ has particularly attracted much attention\cite{Gelman:2005aa,Bruun:2005aa,Massignan:2005aa,Punk:2006aa,Bruun:2007aa,Rupak:2007aa,Taylor:2010aa,Enss:2011aa,Guo:2011aa,Guo:2011ab,LeClair:2011aa,Salasnich:2011aa,Bruun:2012aa,Enss:2012ab,Goldberger:2012aa,Wlazlowski:2012aa,Chafin:2013aa,Romatschke:2013aa,Wlazlowski:2013aa,Bluhm:2014aa,Kryjevski:2014aa,Kikuchi:2016ab,Kagamihara:2017aa,Samanta:2017aa,Cai:2018aa}. One reason is that transport coefficients are related to the particle mean-free-path originating from a pairing interaction, so that they involve useful information about collisional properties of the system in the BCS-BEC crossover region\cite{Gelman:2005aa}. Another reason is the stimulation by the conjecture by Kovtun, Son, and Starinets (KSS)\cite{Kovtun:2005aa}, stating that the ratio of $\eta$ to the entropy density $s$ has the lower bound as,
\begin{equation}
\frac{\eta}{s} \geq \frac{\hbar}{4\pi k_{\rm B}}.
\label{eq.1}
\end{equation}
This was originally proposed in relativistic quantum field theories at finite temperature and zero chemical potential; however, because Eq. (\ref{eq.1}) does not involve the speed of light, KSS speculated that it may be valid for the non-relativistic case, at least for a single-component gas with spin 0 or 1/2. Although some formal counter-examples are known for this KSS conjecture\cite{Cohen:2007aa,Son:2008aa,Brigante:2008aa,Kats:2009aa,Buchel:2009aa,Rebhan:2012aa}, the ratio $\eta/s$ is still considered as a useful quantity for the study of interaction effects on fluid properties. Indeed, Ref. \cite{Schafer:2009ab} evaluated $\eta/s$ for the some quantum fluids from experimental data: (i) liquid $^4$He: $\eta/s\gesim 8.8$, (ii) unitary $^6$Li Fermi gas: $\eta/s\gesim 6.3$, and (iii) quark-gluon plasma: $\eta/s\gesim 5.0$, in unit of $\hbar/(4\pi k_{\rm B})$. For comparison, it is useful to note that $\eta/s\simeq 380\times\hbar/(4\pi k_{\rm B})$ in water under the normal condition\cite{Kovtun:2005aa}. Thus, the results in the above-mentioned quantum fluids are very close to the so-called KSS bound $\eta/s=\hbar/(4\pi k_{\rm B})$. 
\par
From the viewpoint of BCS-BEC crossover physics, because the shear viscosity diverges in an ideal gas, the above result, $\eta/s\gesim 6.3\times \hbar/(4\pi k_{\rm B})$,  obtained in a unitary $^6$Li Fermi gas\cite{Schafer:2009ab} implies that the shear viscosity becomes the smallest around the unitary limit. (Note that the system is reduced to an ideal gas in both the BCS and BEC limits.) Regarding this, it has recently been observed in a $^6$Li Fermi gas\cite{Elliott:2014ab} that, although the shear viscosity $\eta$ becomes small in the unitary regime, the minimum value is obtained, not at the unitarity, but slightly in the strong-coupling BEC side. This makes us expect that the minimum value of $\eta/s$ is also obtained (slightly) away from the unitarity limit.
\par
Motivated by these, in this paper, we theoretically investigate the shear viscosity $\eta$ in an ultracold Fermi gas in the BCS-BEC crossover region. Including strong pairing fluctuations associated with a Feshbach-induced tunable pairing interaction within the framework of the self-consistent $T$-matrix approximation (SCTMA)\cite{Haussmann:1993aa,Haussmann:1994aa,Haussmann:2007aa,Enss:2011aa}, we calculate this transport coefficient, by using the linear response theory. In the first step toward the assessment of the KSS conjecture, this paper focuses on the shear viscosity $\eta$. The ratio $\eta/s$ will separately be discussed in our subsequent paper. We clarify how $\eta$ behaves in the normal state above $T_{\rm c}$ in the BCS-BEC crossover region. We also examine where $\eta$ becomes minimum in the unitary regime, comparing the recent experiment on a $^6$Li Fermi gas\cite{Elliott:2014ab}.
\par
This paper is organized as follows. In Sec. \ref{sec:formulation}, we explain our formulation. The calculated shear viscosity in the BCS-BEC region is shown in Sec. \ref{sec:result}. We also compare our results with the recent experiments on $^6$Li Fermi gases in this section. In the followings, we set $\hbar=k_{\rm B} = 1$, and the system volume is taken to be unity (except in appendix \ref{app:derivation_Pi}), for simplicity.
\par
\par
\section{Formulation}
\label{sec:formulation}
\par
We consider a two-component unpolarized Fermi gas, described by the ordinary BCS Hamiltonian,
\begin{align}
H=\sum_{\bm{p},\sigma} \xi_{\bm{p}} c^{\dag}_{\bm{p},\sigma} c_{\bm{p},\sigma}
-U\sum_{\bm{p},\bm{p}',\bm{q}} c^{\dag}_{\bm{p}+\bm{q},\uparrow} c^{\dag}_{\bm{p}'-\bm{q},\downarrow} c_{\bm{p}',\downarrow} c_{\bm{p},\uparrow},
\label{eq.1.0}
\end{align}
where $c_{\bm{p},\sigma}^\dagger$ is the creation operator of a Fermi atom with pseudospin $\sigma=\uparrow,\downarrow$, describing two atomic hyperfine states. $\xi_{\bm p}=\varepsilon_{\bm p}-\mu=p^2/(2m)-\mu$ is the kinetic energy of a Fermi atom, measured from the Fermi chemical potential $\mu$, where $m$ is an atomic mass. $-U~(<0)$ is a pairing interaction, which is assumed to be tunable by the Feshbach-resonance technique\cite{Chin:2010aa}. The interaction strength is conveniently measured in terms of the $s$-wave scattering length $\as$, which is related to the bare interaction $-U$ as
\begin{equation}
\frac{4\pi a_{s}}{m}=
\frac{-U}{1-U\sum_{\bm p}^{p_{\rm c}}\frac{1}{2\varepsilon_{\bm p}}},
\label{eq_renormalization}
\end{equation}
where $p_{\rm c}$ is a momentum cutoff. In this scale, the weak-coupling BCS regime and the strong-coupling BEC regime are, respectively, characterized as $(k_{\rm F}a_{s})\lesssim -1$, and $(k_{\rm F}a_{s})^{-1}\gtrsim 1$ (where $k_{\rm F}$ is the Fermi wave-length). The region, $-1\lesssim (k_{\rm F}a_{s})^{-1}\lesssim 1$, is referred to as the BCS-BEC crossover regime.
\par
\begin{figure}[t]
\begin{center}
\includegraphics[width=7cm]{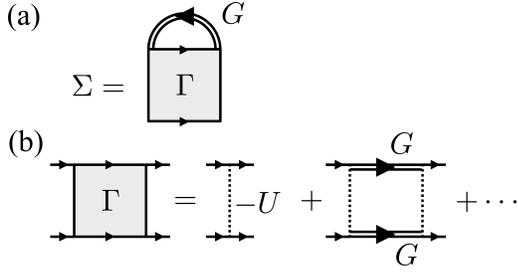}
\end{center}
\caption{(a) Self-energy $\Sigma({\bm p},i\omega_n)$, and (b) particle-particle scattering matrix $\Gamma(\bm{q},i\nu_m)$ in SCTMA. The double solid line is the dressed Green's function $G$ in Eq. (\ref{eq_dressd_G_def}). The dashed line denotes the pairing interaction $-U~(<0)$.}
\label{fig1}
\end{figure}
\par
Strong-coupling corrections to single-particle excitations can be described by the self-energy $\Sigma({\bm p},i\omega_n)$ in the single-particle thermal Green's function,
\begin{equation}
G(\bm{p},i\omega_n) = \frac{1}{i\omega_n - \xi_{\bm{p}} - \Sigma(\bm{p},i\omega_n)},
\label{eq_dressd_G_def}
\end{equation}
where $\omega_n$ is the fermion Matsubara frequency. In SCTMA\cite{Haussmann:1994aa,Haussmann:1993aa,Haussmann:2007aa,Enss:2011aa}, $\Sigma({\bm p},i\omega_n)$ is obtained from the diagrams in Fig. \ref{fig1}, which gives
\begin{equation}
\Sigma(\bm{p},i\omega_n) = T \sum_{\bm{q},\nu_m} \Gamma(\bm{q},i\nu_m) G(\bm{q} - \bm{p}, i\nu_m - i\omega_n).
\label{eq_self_energy}
\end{equation}
Here, $\nu_m$ is the boson Matsubara frequency, and 
\begin{align}
\Gamma(\bm{q},i\nu_m) 
&= \frac{-U}{1-U\Lambda(\bm{q},i\nu_m)}
\nonumber
\\
&=
\frac{4\pi a_s}{m}
\frac{1}{1+\frac{4\pi a_s}{m}\left[ \Lambda({\bm q},i\nu_m)-\sum_{\bm p} \frac{1}{2\varepsilon_{\bm p}} \right]}
\label{eq.1.1}
\end{align}
is the SCTMA particle-particle scattering matrix, describing fluctuations in the Cooper channel (see also Fig. \ref{fig1}(b)). Here, 
\begin{equation}
\Lambda(\bm{q},i\nu_m) = T \sum_{\bm{p},\omega_n} G(\bm{p},i\omega_n) G(\bm{q}-\bm{p},i\nu_m-i\omega_n),
\label{eq_Pi_def}
\end{equation}
is the pair correlation function.
\par
\begin{figure}[t]
\begin{center}
\includegraphics[width=7cm]{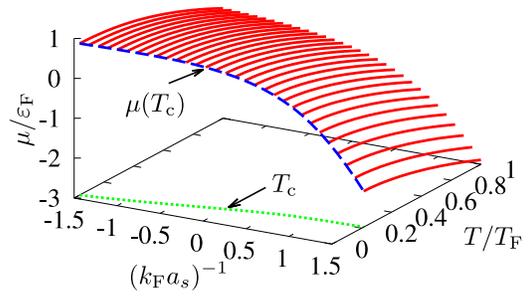}
\end{center}
\caption{SCTMA Fermi chemical potential $\mu$ in the BCS-BEC crossover region above $T_{\rm c}$. $\varepsilon_{\rm F}$ and $T_{\rm F}$ are the Fermi energy and the Fermi temperature, respectively.
}
\label{fig2}
\end{figure}

In this scheme, we determine $T_{\rm c}$ from the Thouless criterion\cite{Thouless:1960aa}, stating that the superfluid instability occurs, when the particle-particle scattering matrix in Eq. (\ref{eq.1.1}) has a pole at ${\bm q}=\nu_m=0$, which gives
\begin{equation}
1=-\frac{4\pi a_s}{m}
\left[
\Lambda(\bm{q}=0,i\nu_m=0)
- 
\sum_{\bm{p}} \frac{1}{2\varepsilon_{\bm{p}}}
\right].
\label{eq.1.2}
\end{equation}
As in the ordinary BCS-BEC crossover theories\cite{Eagles:1969aa,Leggett:1980aa,Nozieres:1985aa,Sa-de-Melo:1993aa,Haussmann:1993aa,Haussmann:1994aa,Pistolesi:1994aa,Holland:2001aa,Timmermans:2001aa,Ohashi:2002aa}, we actually solve the $T_{\rm c}$-equation (\ref{eq.1.2}), together with the equation for the number $N$ of Fermi atoms,
\begin{equation}
N = 2T\sum_{\bm{p},\omega_n} G(\bm{p},i\omega_n),
\label{eq_number_eq_SCTMA}
\end{equation}
to self-consistently determine $T_{\rm c}$ and $\mu(T_{\rm c})$. Above $T_{\rm c}$, we only treat the number equation (\ref{eq_number_eq_SCTMA}), to evaluate $\mu(T)$. We briefly show the self-consistent solution for $\mu(T\ge T_{\rm c})$ in Fig, \ref{fig2}, which is used in calculating $\eta(T)$. 
\par
\par
In the linear response theory\cite{Fetter:2003aa,Rickayzen:2013aa}, the shear viscosity $\eta$ is given by\cite{Kadanoff:1963aa,Luttinger:1964aa,Bruun:2007aa,Taylor:2010aa},
\begin{equation}
\eta = - \lim_{\omega \to 0}
\frac{1}{\omega}{\rm Im}[\Xi(\omega)].
\label{eq.100}
\end{equation}
Here, 
\begin{equation}
\Xi(\omega) = -i 
\int d^3\bm{r} 
\int_0^\infty dt e^{i\omega t}
\langle [ \hat{\Pi}_{x,y}(\bm{r},t), \hat{\Pi}_{x,y}(\bm{0},0) ] \rangle
\label{eq_Xi_definition}
\end{equation}
is the shear-stress response function, where $\hat{\Pi}_{x,y}$ is the $xy$ component of the stress tensor operator. In the present BCS model in Eq. (\ref{eq.1.0}), it has the form,
\begin{align}
\hat{\Pi}_{x,y}(\bm{r}) 
&= \sum_{\sigma} \frac{1}{2m}
\left[ (\nabla_i \psi^{\dag}_{\sigma})(\nabla_j \psi_{\sigma}) 
+ (i\leftrightarrow j)
-\frac{1}{2} \nabla_i \nabla_j (\psi^{\dag}_{\sigma} \psi_{\sigma})
\right],
\label{eq.1.5}
\end{align}
where the field operator $\psi_\sigma({\bm r})$ describes Fermi atoms with pseudospin $\sigma=\uparrow,\downarrow$. For the derivation of Eq (\ref{eq.1.5}), see Appendix \ref{app:derivation_Pi}. 
\par
In this paper, we first evaluate the corresponding thermal response function,
\begin{equation}
\Xi(i\nu_l) = -\int d^3\bm{r} \int_0^{\beta} e^{i\nu_l \tau} \langle T_{\tau} \hat{\Pi}_{x,y}(\bm{r},\tau) \hat{\Pi}_{x,y}(\bm{0},0) \rangle,
\label{eq2-12}
\end{equation}
where $T_{\tau}$ is the imaginary-time-ordered-product and $\nu_l$ is the boson Matsubara frequency. $\Xi(\omega)$ is then obtained as $\Xi(\omega) = \Xi(i\nu_l \to \omega_+=\omega + i\delta)$, where $\delta$ is an infinitesimally small positive number. In this paper, we numerically carry out this analytic continuation by the Pad\'e approximation\cite{Vidberg:1977aa}.
\par
\par
\begin{figure}[t]
\begin{center}
\includegraphics[width=7cm]{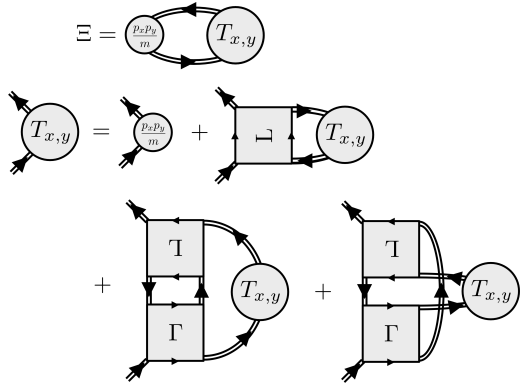}
\end{center}
\caption{Shear-stress response function $\Xi(i\nu_l)$ in SCTMA. The small (large) circle represents the bare (dressed) three point vertex $p_x p_y/m$ ($T_{x,y}$). The double solid line is the dressed Green's function $G$. `MT' and `AL' diagrams give Eqs. (\ref{eq_T_MT_def}) and (\ref{eq_T_AL_def}), respectively.
}
\label{fig3}
\end{figure}
\par
In evaluating the thermal response function in Eq. (\ref{eq2-12}), we need to choose diagrams so as to satisfy the Ward-Takahashi identity\cite{Schrieffer:1999aa,Baym:1961ab,Baym:1962aa,He:2014aa}, which is a required condition in any consistent theory. For the shear-stress response function $\Xi(i\nu_l)$, this identity is derived from the momentum conservation law\cite{Enss:2011aa,He:2014aa}. In SCTMA, it is diagrammatically described in Fig. \ref{fig3}\cite{Enss:2011aa}, giving
\begin{align}
\Xi(i\nu_l)
&= 2 T \sum_{\bp,\omega_n} \frac{p_x p_y}{m} G(\bm{p},i\omega_n) 
T_{x,y}(\bm{p},i\omega_n,i\omega_n+i\nu_l)
\nonumber
\\
&\hskip10mm
\times
G(\bm{p},i\omega_n+i\nu_l).
\label{eq_Xi_by_vertex}
\end{align}
Here, the three point vertex function $T_{x,y}=p_xp_y/m+T_{x,y}^{\rm MT}+T_{x,y}^{\rm AL}$ consists of the bare term $p_xp_y/m$, Maki-Thompson (MT) term, 
\begin{align}
&\hskip-3mm
T_{x,y}^{\rm MT}(\bm{p},i\omega_n,i\omega_n+i\nu_l)
\nonumber
\\
&\hskip-4mm
= T \sum_{\bq,\nu_m} \Gamma(\bq,\inu) \tilde{T}_{x,y}(\bq-\bp,\inu-\iom-i\nu_l,\inu-\iom),
\label{eq_T_MT_def}
\end{align}
as well as the Aslamazov-Larkin (AL) term,
\begin{align}
&T_{x,y}^{\rm AL}(\bm{p},i\omega_n,i\omega_n+i\nu_l)
\nonumber
\\
&
= -2T \sum_{\bq,\nu_m} \tilde{S}_{x,y}(\bq,\inu,\inu+i\nu_l) G(\bq-\bp,\inu-\iom).
\label{eq_T_AL_def}
\end{align}
Here, ${\tilde T}_{x,y}$ and ${\tilde S}_{x,y}$ are, respectively, given as follows:
\begin{align}
&\tilde{T}_{x,y}(\bp,\iom,\iom+i\nu_l)
\nonumber
\\
&\hskip5mm= G(\bp,\iom) T_{x,y}(\bp,\iom,\iom+i\nu_l) G(\bp,\iom+i\nu_l),
\end{align}
\begin{align}
&\tilde{S}_{x,y}(\bq,\inu,\inu+i\nu_l) = \Gamma(\bq,\inu)\Gamma(\bq,\inu+i\nu_l)
\nonumber
\\
&\hskip5mm \times T\sum_{\bp,\omega_n} G(\bq-\bp,\inu-\iom) \tilde{T}_{x,y}(\bp,\iom,\iom+i\nu_l).
\label{eq_S_def}
\end{align}
\par
\begin{figure}[t]
\begin{center}
\includegraphics[width=8cm]{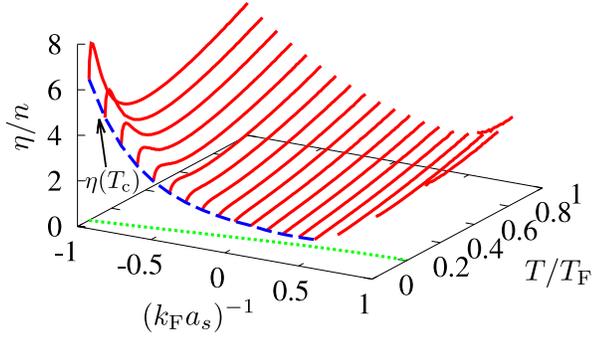}
\end{center}
\caption{Calculated shear viscosity $\eta(T)$ in the BCS-BEC crossover regime of an ultracold Fermi gas above $T_{\rm c}$. The dotted line shows $\eta(T_{\rm c})$. $n$ is the number density of Fermi atoms. The Pad\'e approximation did not work well near $T_{\rm c}$ in the BEC regime, so that the figure does not the result there.
}
\label{fig4}
\end{figure}

Before ending this section, we comment on our numerical calculations to obtain $\eta(T)$. In computing the shear viscosity, we have sometimes met the difficulty that the Pad\'e approximation unphysically gives negative $\eta$ or positive but abnormally large/small $\eta$. We have also found that this problem depends on the detailed choice of momentum cutoff in numerically evaluating $\Xi(i\nu_l)$ in Eq. (\ref{eq_Xi_by_vertex}), as well as on the number of Matsubara frequencies in executing the numerical analytic continuation by the Pad\'e approximation. At this stage, we have no idea to completely overcome this problem. Thus, we have employed the following prescription in this paper: (1) We first calculate $\Xi(i\nu_l)$ by introducing various values of the momentum cutoff $k_{\rm c}$ ($10k_{\rm F}\lesssim k_{\rm c}\lesssim 60k_{\rm F}$) to the momentum summation in Eq. (\ref{eq_Xi_by_vertex}) (which is nothing to do with $p_{\rm c}$ in Eq. (\ref{eq_renormalization})). (2) For each result, we next numerically execute the analytic continuation by the Pad\'e approximation\cite{Vidberg:1977aa}, retaining $50\sim 100$ Matsubara frequencies $\nu_l$. (3) For these data set, we remove clearly unphysical negative data, and then remove the highest and lowest 10\% of data to avoid influence of abnormal results. (4) For the remaining data, we evaluate the averaged value $\bar{\eta}$, as well as the standard derivation ${\bar \sigma}$. When $|{\bar \sigma} /\bar{\eta}|$ is less than 0.1, we plot it in Fig. \ref{fig4}. Otherwise, we judge that the result is not reliable, not to plot it in Fig. \ref{fig4}. The latter situation occurs near $T_{\rm c}$ in the BEC regime, so that $\eta$ is not shown there in this figure.
\par
\par
\begin{figure}[t]
\begin{center}
\includegraphics[width=7cm]{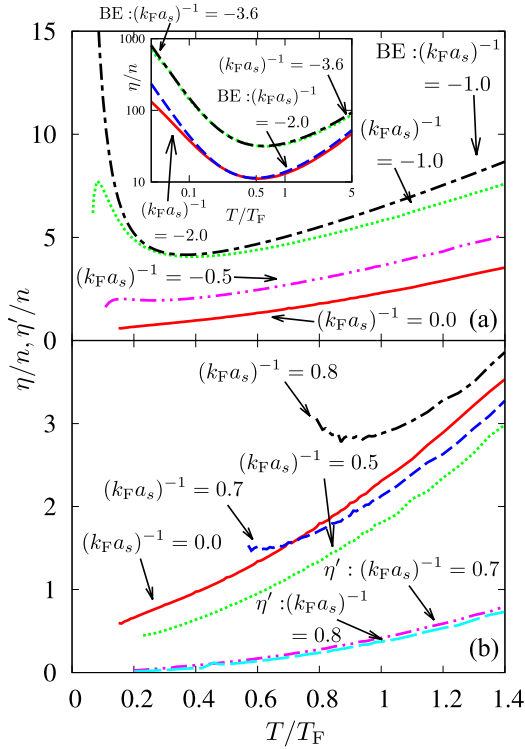}
\end{center}
\caption{Shear viscosity $\eta$ as a function temperature, at several interaction strengths. (a) BCS side ($(k_{\rm F}a_s)^{-1}\lesssim 0$). (b) BEC side ($(k_{\rm F}a_s)^{-1}\gesim 0$). $\eta'$ is the result in the case when the vertex corrections (MT and AL diagrams in Fig. \ref{fig3} are removed. The absence of the result near $T_{\rm c}$ in panel (b) is due to the computational problem explained in Sec. \ref{sec:formulation}. `BE' is the result by using the Boltzmann equation (\ref{eq.b5}). In the inset, the logarithmic scale is used.
}
\label{fig5}
\end{figure}
\par
\section{Shear viscosity $\eta$ in the BCS-BEC crossover regime of an ultracold Fermi gas}
\label{sec:result}
\par
Figure \ref{fig4} shows the shear viscosity $\eta(T)$ in the BCS-BEC crossover regime of an ultracold Fermi gas above $T_{\rm c}$. When $(k_{\rm F}a_s)^{-1}\lesssim -0.5$ in the weak-coupling BCS side, we find that, with decreasing the temperature from the Fermi temperature $T_{\rm F}$, $\eta$ exhibits a dip structure. Deep inside the BCS regime ($(k_{\rm F}a_s)^{-1}\lesssim -2$), Fig. \ref{fig5}(a), as well as the inset in this panel, show that this non-monotonic temperature dependence is well reproduced by the Boltzmann equation. As summarized in Appendix \ref{app:Boltzmann}, in the classical regime ($T\gg T_{\rm F}$), the Boltzmann-equation approach gives $\eta(T)\propto T^{3/2}$ ($T\gg 1/(ma_s^2)$) and $\eta(T)\propto T^{1/2}$ ($T\ll 1/(ma_s^2)$)\cite{Massignan:2005aa,Bruun:2005aa}, both of which {\it decreases} with decreasing the temperature. In the Fermi degenerate regime ($T\ll T_{\rm F}$), this approach gives $\eta(T)\propto T^{-2}$, which {\it increases} as the temperature decreases. These explain the behavior of $\eta(T)$ above and below the dip temperature ($\equiv T^{\rm BCS}_{\rm dip}$), respectively. Thus, this dip structure is nothing to do with the BCS-BEC crossover phenomenon, but is a phenomenon which can be explained within the standard Boltzmann equation ignoring pairing fluctuations\cite{note1}.
\par
\begin{figure}[t]
\begin{center}
\includegraphics[width=7cm]{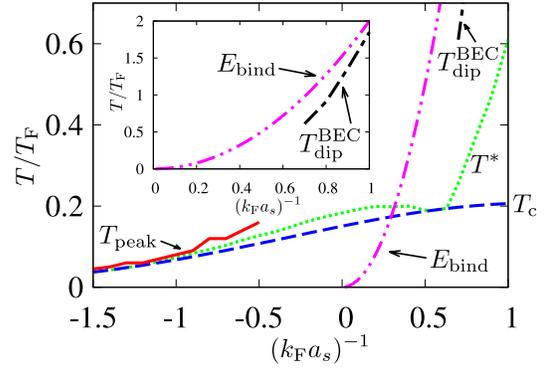}
\end{center}
\caption{Comparison of the temperature $T_{\rm peak}$, at which $\eta(T)$ exhibits a peak structure, and the pseudogap temperature $T^*$, below which the single-particle density of states $\rho(\omega)$ in Eq. (\ref{eq.dos}) has a dip structure around the Fermi level. $T_{\rm peak}$ is obtained in the weak-coupling regime when $-1.5 \lesssim (k_{\rm F}a_{s})^{-1}\lesssim -0.5$. This figure also compares the dip temperature $T_{\rm dip}^{\rm BEC}$, at which $\eta(T)$ exhibits a dip structure in the BEC side, with the binding energy $E_{\rm bind}=1/(ma_s^2)$ of a two-body bound molecule.}
\label{fig6}
\end{figure}

In addition to the dip, we also see in Fig. \ref{fig4} a peak structure near $T_{\rm c}$ when $(k_{\rm F}a_s)^{-1}\lesssim -0.5$. As shown in Fig. \ref{fig5}(a), the Boltzmann equation cannot explain this anomaly, indicating that it is a many-body phenomenon originating from pairing fluctuations enhanced near $T_{\rm c}$ (that are ignored in the Boltzmann equation). Indeed, Fig. \ref{fig6} shows that the peak temperature $T_{\rm peak}$ is close to the so-called pseudogap temperature $T^*$\cite{Hanai}, below which the single-particle density of states,
\begin{align}
\rho(\omega) = - \frac{1}{ \pi}
\sum_{\bp} {\rm Im}[G(\bp,\iom \to \omega+i\delta)],
\label{eq.dos}
\end{align}
has a BCS-state like dip structure around the Fermi level $\omega=0$, due to strong pairing fluctuations (or the formation of preformed Cooper pairs). 
\par
To simply understand the role of pairing fluctuations around $T_{\rm peak}$, we conveniently employ the expression for the shear viscosity $\eta\sim nl_{\rm mfp}{\bar p}$\cite{Pitaevskii:1981aa}, where $n$ is the number density of Fermi atoms, $l_{\rm mfp}$ the mean free path, and ${\bar p}$ is the averaged particle momentum. When the effective interaction between Fermi atoms described by the particle-particle scattering matrix $\Gamma$ in Eq. (\ref{eq.1.1}) is temperature-independent far above $T_{\rm c}$, the quasi-particle lifetime $\tau$ in the mean free path $l_{\rm mfp}\sim (k_{\rm F}/m)\tau$ behaves as, symbolically, $\tau^{-1}\sim |\Gamma|^2 T^2\propto T^2$ when $T\ll T_{\rm F}$, as in the ordinary Fermi liquid theory. The resulting $\eta(T)\propto T^{-2}$ explains the temperature dependence of $\eta(T)$ in the region $T_{\rm peak}\le T\le T^{\rm BCS}_{\rm dip}$. Near $T_{\rm c}$, on the other hand, the particle-particle scattering matrix $\Gamma(0,0)$ is enhanced, to eventually diverge at $T_{\rm c}$ (Thouless criterion). This enhancement of $\Gamma(0,0)$ with decreasing the temperature shortens the quasi-particle lifetime $\tau$, which leads to the decrease of $\eta(T)\propto \tau\sim[|\Gamma|T]^{-2}$, giving the peak structure in Fig. \ref{fig4}. 
\par
\par
We briefly note that low-energy properties of a repulsively interacting Fermi gas at low temperatures are known to be well described by the Fermi liquid theory with the quasi-particle lifetime $\tau\propto T^{-2}$. Regarding this, although the present interaction is {\it attractive}, one still sees the ``Fermi liquid {\it like} behavior'', $\eta(T)\propto\tau\propto T^{-2}$, in the intermediate temperature region, $T_{\rm peak}\lesssim T \lesssim T_{\rm dip}^{\rm BCS}$.
\par
\begin{figure}[t]
\begin{center}
\includegraphics[width=7cm]{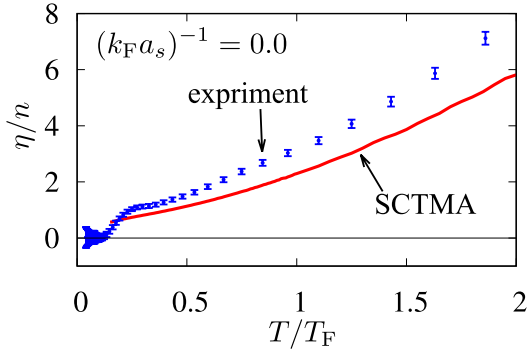}
\end{center}
\caption{Comparison of SCTMA shear viscosity $\eta(T)$ and the recent experiment on a $^6$Li unitary Fermi gas\cite{Joseph:2015aa}.
}
\label{fig7}
\end{figure}

Figures \ref{fig4} and \ref{fig5} show that the above-mentioned dip-peak structure disappears in the unitary regime (-$0.5\lesssim (k_{\rm F}a_s)^{-1}\lesssim 0.5$), because of the smearing of the Fermi surface, as well as the enhancement of fluctuations in the Cooper channel, by the strong pairing interaction between Fermi atoms. The resulting shear viscosity $\eta(T)$ monotonically decreases with decreasing the temperature down to $T_{\rm c}$. We point out that such behavior has recently been observed in a $^6$Li unitary Fermi gas\cite{Joseph:2015aa}; our result semi-quantitatively agrees with this experiment, as shown in Fig. \ref{fig7}. We briefly note that this behavior of $\eta(T)$ has also theoretically been obtained in Ref. \cite{Enss:2011aa}.
\par
As one passes through the unitarity regime to enter the strong-coupling BEC regime ($(k_{\rm F}a_s)^{-1}\gesim 0.7$), the shear viscosity $\eta(T)$ again becomes large, as shown in Figs. \ref{fig4} and \ref{fig5}(b). In addition, although we cannot completely obtain $\eta$ near $T_{\rm c}$ in the BEC regime because of the computational problem mentioned in Sec. \ref{sec:formulation}, we still see in Fig. \ref{fig5}(b) that $\eta(T)$ again exhibits a non-monotonic behavior, when $(k_{\rm F}a_s)^{-1}\gesim 0.7$. Plotting the dip temperature ($\equiv T_{\rm dip}^{\rm BEC}$) in this regime, we find that it is close to the binding energy $E_{\rm bind}=1/(ma_s^2)$ of a two-body bound state (see Fig. \ref{fig6}). This implies that the dip structure in the BEC regime is associated with the start of the formation of two-body bound molecules below $T_{\rm dip}^{\rm BEC}$, overwhelming thermal dissociation.
\par
\par
\begin{figure}[t]
\begin{center}
\includegraphics[width=7cm]{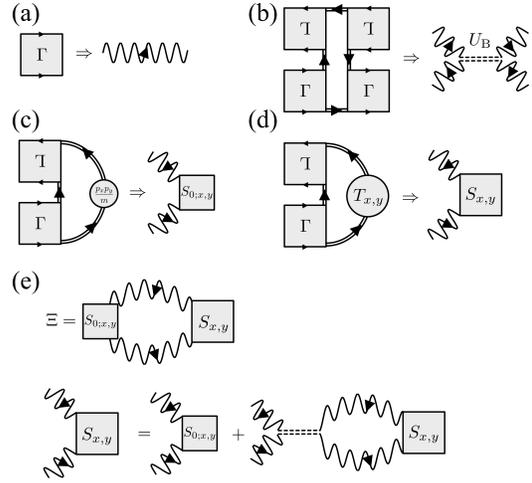}
\end{center}
\caption{Bosonic interpretation of the shear-stress response function $\Xi(i\nu_l)$ in the strong-coupling BEC regime. In this regime, the particle particle scattering matrix $\Gamma$ may be viewed as a Bose propagator (wavy line in (a)). Then, introducing an effective interaction $U_{\rm B}$ between bosons mediated by four unpaired fermions (b), as well as molecular three-point vertex functions $S_{0;x,y}$ (c) and $S_{x,y}$ (d), we obtain from the AL diagrams in Fig. \ref{fig3} the Bose shear-stress response function in (e).\cite{note2}}
\label{fig8}
\end{figure}

To explain how Bose-molecules affect the shear viscosity $\eta$ in the BEC regime, it is convenient to represent the shear-stress response function $\Xi(i\nu_l)$ with respect to the Bose degrees of freedom: In the extreme BEC limit ($(k_{\rm F}a_s)^{-1}\gg 1$), the particle-particle scattering matrix $\Gamma({\bm q},i\nu_l)$ is known to describe tightly bound molecular bosons. To simply see this, approximately evaluating the pair correlation function $\Lambda({\bm q},i\nu_m)$ in Eq. (\ref{eq_Pi_def}) by replacing the SCTMA dressed Green's function with the bare one, $G_0({\bm p},i\omega_n)=[i\omega_n-\xi_{\bm p}]^{-1}$, one finds\cite{Haussmann:1993aa}, in the BEC limit,
\begin{equation}
\Gamma(\bm{q},i\nu_m) \simeq \frac{8\pi}{m^2 \as} \frac{1}{i\nu_m - \xi_{\bm q}^{\rm B}}.
\end{equation}
Here, $\xi_{\bm q}^{\rm B}={\bm q}^2/(4m)-\mu_{\rm B}$ is the molecular kinetic energy, measured from the Bose chemical potential $\mu_{\rm B} = 2 \mu + E_{\rm bind}$. Then, (1) regarding the particle-particle scattering matrix $\Gamma$ appearing in the diagrams of $\Xi(i\nu_l)$ in Fig. \ref{fig3} as the Bose Green's function (wavy line in Fig. \ref{fig8}(a)), and (2) further introducing an effective Bose-Bose interaction $U_{\rm B}$ as Fig. \ref{fig8}(b), as well as an bare and dressed molecular three-point vertex functions $S_{0;x,y}$ and $S_{x,y}$ as Figs. \ref{fig8}(c) and (d), respectively, we find that the AL-type diagrams in Fig. \ref{fig3} involves the contribution that can be written in the form of the Bose shear-stress response function, as shown in Fig. \ref{fig8}(e). 
\par
\par
In SCTMA, the effective molecular interaction $U_{\rm B}$, that are mediated by four unpaired Fermi atoms (see Fig. \ref{fig8}(b)), is repulsive and is given by\cite{Haussmann:1993aa} 
\begin{equation}
U_{\rm B}= \frac{4\pi a_{\rm B}}{m_{\rm B}},
\label{eq.intB}
\end{equation}
where $m_{\rm B}=2m$ is a molecular mass and $a_{\rm B}=2\as>0$ is the $s$-wave molecular scattering. In the BEC regime, because most Fermi atoms form tightly bound molecules described by the wavy line in Fig. \ref{fig8}, the shear viscosity is considered to be dominated by the molecular shear-stress response function in Fig. \ref{fig8}(e). This contribution becomes large with decreasing the molecular interaction strength $U_{\rm B}\propto 2a_s\to 0$ with approaching the BEC limit $(k_{\rm F}a_s)^{-1}\to +\infty$. (Note that the shear viscosity diverges in an ideal Bose gas.)
\par
\begin{figure}[t]
\begin{center}
\includegraphics[width=7cm]{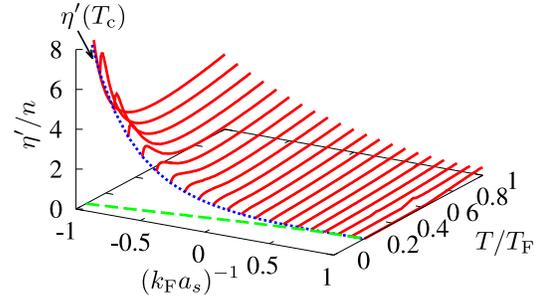}
\end{center}
\caption{Calculated shear viscosity $\eta'(T)$ {\it without} the MT-type and AL-type vertex corrections in Fig. \ref{fig3}. The dotted line shows the result at $T_{\rm c}$.}
\label{fig9}
\end{figure}

To support the above diagrammatic discussion, we show in Fig. \ref{fig9} the result ($\equiv \eta'$) in the case when we only retain the first term in Fig. \ref{fig3} and ignore all the MT and AL vertex corrections. In this figure, we see that $\eta'$ simply decreases as one approaches the BEC regime, because it does not involve molecular contribution to the shear viscosity. In addition, as shown in Fig. \ref{fig5}(b), $\eta'$ monotonically decreases with decreasing the temperature when ($(k_{\rm F}a_s)^{-1}\ge 0.7$), being in contrast to the up-turn behavior of $\eta(T)$ including vertex corrections. 
\par
When we apply the simple expression for the shear viscosity $\eta \sim nl_{\rm mfp}{\bar p}$ to the molecular Bose gas in the BEC regime, the mean-free path $l_{\rm mfp}$ is expected to be closely related to the molecular lifetime $\tau_{\rm B}$. Regarding this, we note that, although SCTMA includes the molecular interaction $U_{\rm B}$ in Eq. (\ref{eq.intB}), this strong-coupling theory involves {\it effects} of this repulsive interaction only in the mean-field level\cite{Haussmann:1993aa}. That is, the molecular lifetime $\tau_{\rm B}$ coming from $U_{\rm B}$ is completely ignored in SCTMA. However, because $\eta \sim nl_{\rm mfp}{\bar p}$ diverges when $l_{\rm mfp}\propto\tau_{\rm B}=\infty$, the converging $\eta(T)$ in Fig. \ref{fig5}(b) implies the presence of another scattering process contributing to the molecular lifetime $\tau_{\rm B}$.
\par
As the origin of $\tau_{\rm B}$ in SCTMA, we note that, because the particle-particle scattering matrix $\Gamma({\bm q},i\nu_m)$ in Eq. (\ref{eq.1.1}) is reduced to the single-particle Bose Green's function in the BEC regime, the lifetime of the boson should come from the pair correlation function, ${\rm Im}[\Lambda({\bm q},i\nu_m\to\omega_+)]~(\equiv\gamma({\bm q},\omega))$. To confirm this, we evaluate this molecular damping $\gamma({\bm q},\omega)$ by simply approximating the self-energy $\Sigma({\bm p},i\omega_n)$ in Eq. (\ref{eq_self_energy}) involved in the dressed Green's function $G$ to\cite{Tsuchiya:2009aa}
\begin{align}
\Sigma({\bm p},i\omega_n)
&\simeq
T\sum_{{\bm q},\nu_m}
\Gamma({\bm q},i\nu_m)G_0({\bm q}-{\bm p},i\nu_m-i\omega_n)
\nonumber
\\
&\simeq
G_0(-{\bm p},-i\omega_n)\times
T\sum_{{\bm q},\nu_m}
\Gamma({\bm q},i\nu_m)
\nonumber
\\
&\equiv-\Delta_{\rm pg}^2
G_0(-{\bm p},-i\omega_n).
\label{app.self}
\end{align}
Here, we have approximately set ${\bm q}=\nu_m=0$ in the bare Green's function $G_0$, using the fact that $\Gamma(0,0)$ is enhanced near $T_{\rm c}$. $\Delta_{\rm pg}$ is the so-called pseudogap parameter\cite{Chen:2009aa,Tsuchiya:2009aa}, physically describing effects of pairing fluctuations on single-particle excitations. The resulting dressed Green's function $G$ formally has the same form as the diagonal component of the BCS Green's function\cite{Fetter:2003aa,Rickayzen:2013aa} as,
\begin{equation}
G({\bm p},i\omega_n)=
-\frac{i\omega_n+\xi_{\bm p}}{\omega_n^2+\xi_{\bm p}^2+\Delta_{\rm pg}^2}.
\label{eq.pg}
\end{equation}
Substituting Eq. (\ref{eq.pg}) into Eq. (\ref{eq_Pi_def}), as well as executing the $\omega_n$-summation, we obtain, after the analytic continuation $i\nu_m\to\omega_+$,
\begin{align}
\gamma({\bm q},\omega\ge 0)
&=\frac{\pi}{4}
\sum_{\bm p}
\left[
1+ \frac{\xi_{{\bm q}/2+{\bm p}}}{E_{{\bm q}/2+{\bm p}}}
\right]
\left[
1 + \frac{\xi_{{\bm q}/2-{\bm p}}}{E_{{\bm q}/2-{\bm p}}}
\right]
\nonumber
\\
&\hskip-15mm\times
[1-f_{\rm F}(E_{{\bm q}/2+{\bm p}})-f_{\rm F}(E_{{\bm q}/2-{\bm p}})
]
\delta(\omega-[E_{{\bm q}/2+{\bm p}}+E_{{\bm q}/2-{\bm p}}])
\nonumber
\\
&-
\frac{\pi}{2}
\sum_{\bm p}
\left[
1+\frac{\xi_{{\bm q}/2+{\bm p}}}{E_{{\bm q}/2+{\bm p}}}
\right]
\left[
1-\frac{\xi_{{\bm q}/2-{\bm p}}}{E_{{\bm q}/2-{\bm p}}}
\right]
\nonumber
\\
&\hskip-15mm\times
[f_{\rm F}(E_{{\bm q}/2+{\bm p}})-f_{\rm F}(E_{{\bm q}/2-{\bm p}})]
\delta(\omega-[E_{{\bm q}/2+{\bm p}}-E_{{\bm q}/2-{\bm p}}]),
\label{eq.damp}
\end{align}
where $E_{\bm p}=\sqrt{\xi_{\bm p}^2+\Delta_{\rm pg}^2}$ and $f_{\rm F}(x)=1/[\exp(x/T)+1]$ is the equilibrium Fermi distribution function.
 In Eq. (\ref{eq.damp}), the first term comes from the so-called inter-band excitation\cite{OhashiTakada}. Since this excitation is accompanied by pair-breaking, it has the excitation threshold, $\omega_{\rm th}={\rm Min}[E_{{\bm q}/2+{\bm p}}+E_{{\bm q}/2-{\bm p}}]=2|\Delta_{\rm pg}|$. As a result, this term does not contribute to the damping $\gamma({\bm q},\omega)$, as far as we consider low-energy excitations $\omega\le \omega_{\rm th}$. 
\par
The second term in Eq. (\ref{eq.damp}) is associated with the intra-band excitation\cite{OhashiTakada}. Since this is an excitation of an unpaired Fermi atom which has already been excited thermally, it has no excitation threshold. Thus, the molecular damping $\gamma({\bm q},\omega)$ is dominated by this process. Noting that $\mu\to-E_{\rm bind}/2=-1/(2ma_s^2)\ll-\varepsilon_{\rm F}$\cite{Leggett:1980aa,Nozieres:1985aa,Sa-de-Melo:1993aa,Haussmann:1993aa,Haussmann:1994aa} and $E_{\bm p}\simeq \varepsilon_{\bm p}+E_{\rm bind}/2$ in the strong-coupling BEC regime, one obtains, in the low-energy and low-momentum regime,
\begin{align}
\gamma(\bm{q},\omega)
&\simeq
\frac{m^2\Delta_{\rm pg}^2}{8\pi |\mu|^2}
\left( \frac{\omega}{q} \right)
e^{-\frac{m}{2T} \left(\frac{\omega}{q}\right)^2}
e^{-\frac{E_{\rm bind}}{2T}}.
\label{eq.damp2}
\end{align}
Even in the BEC regime, the thermal dissociation of bound molecules gives unpaired Fermi atoms to some extent, which is reflected by the factor $e^{-E_{\rm bind}/(2T)}~(\ll 1)$ in Eq. (\ref{eq.damp2}). This small factor describes the longevity of molecular bosons in this regime:
\begin{equation}
\tau_{\rm B}({\bm q},\omega)
=
\frac{m^2 a_s}{8\pi} \frac{1}{\gamma({\bm q},\omega)}
=
\frac{a_sE_{\rm bind}^2}{4\Delta_{\rm pg}^2}
\left( \frac{q}{\omega} \right)
e^{\frac{m}{2T} \left(\frac{\omega}{q}\right)^2}
e^{\frac{E_{\rm bind}}{2T}}.
\label{eq.tau_B}
\end{equation}
\par
\par
In the strong-coupling BEC regime where most Fermi atoms form tightly bound molecules, $\eta$ is dominated by the molecular contribution in Fig. \ref{fig8}(e).
To see how the long-lived molecules affect $\eta(T)$ in this regime, we consider a model Bose gas with $\tau_{\rm B}=\lambda e^{E_{\rm bind}/(2T)}$, ignoring the energy- and momentum-dependence that Eq. (\ref{eq.tau_B}) possesses. In this simple model, we obtain (For the derivation, see Appendix \ref{app:derivation_eq.eta_BEC}.),
\begin{equation}
\eta =
-\frac{\lambda}{2}
e^{\frac{E_{\rm bind}}{2T}}
\sum_{\bm{q}} 
\left( \frac{q_x q_y}{2m} \right)^2
\left( \frac{\partial n_{\rm B}(\xi_{\bm q}^{\rm B})}{\partial \xi_{\bm q}^{\rm B}} \right),
\label{eq.eta_BEC}
\end{equation}
where $n_{\rm B}(x) =1/[\exp(x/T)-1]$ is the Bose distribution function. We find from Eq. (\ref{eq.eta_BEC}), as well as the result $T_{\rm dip}^{\rm BEC}\sim E_{\rm bind}$ shown in the inset in Fig. \ref{fig6}, that the enhancement of $\eta(T)$ below $T_{\rm dip}^{\rm BEC}$ in Fig. \ref{fig5}(b) originates from the appearance of very long-lived molecular bosons.
\par
As mentioned previously, although SCTMA ignores molecular scatterings by the effective interaction $U_{\rm B}$ in Eq. (\ref{eq.tau_B}), these should actually contribute the molecular lifetime $\tau_{\rm B}$. Although inclusion of this effect remains as our future problem, we speculate that it would decrease the magnitude of $\eta(T)$ in the BEC regime, especially below $T_{\rm dip}^{\rm BEC}$, compared to the present SCTMA result.
\par
\begin{figure}[t]
\begin{center}
\includegraphics[width=7cm]{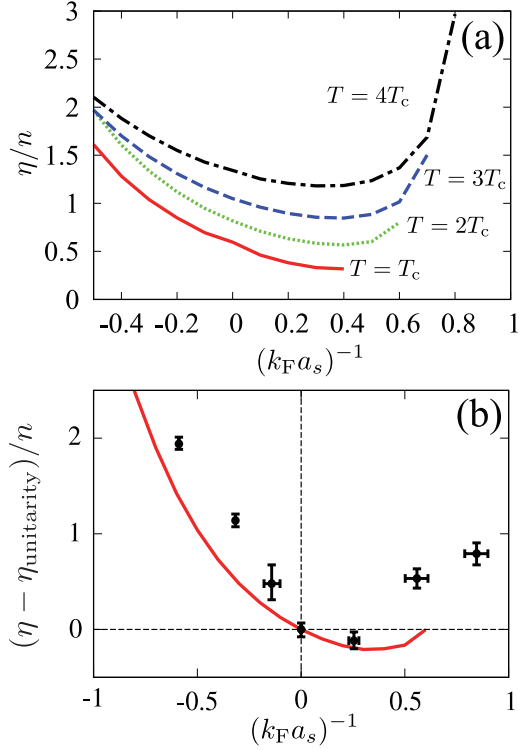}
\end{center}
\caption{(a) SCTMA shear viscosity $\eta(T)$ as a function of the interaction strength above $T_{\rm c}$. (b) Comparison of the SCTMA result (solid line) with the experimental result in a $^6$Li Fermi gas (filled circles)\cite{Elliott:2014ab}.
 The shear viscosity is measured from the value $\eta_{\rm unitarity}$ in the unitarity limit. We note that the experimental data in panel (b) are the trap-averaged ones, under the condition in Eq. (\ref{eq.cond}). To model this in our uniform theory, we have imposed the condition $3P/N = \ef$ in calculating $\eta$ at each interaction strength (where $P$ is the pressure).
}
\label{fig10}
\end{figure}
\par
Figure \ref{fig10}(a) shows $\eta$ as a function of the interaction strength. In this figure, we find that the minimum of $\eta$ is obtained, not in the unitarity limit, but slightly in the BEC side, being consistent with the recent experiment on a $^6$Li Fermi gas\cite{Elliott:2014ab}. However, when we directly compare our result with this experiment, we should note that the experiment was done in a trap potential, under the condition,
\begin{align}
\frac{3}{N} \int d\bm{r} p(\bm{r}) = \varepsilon_{\rm F},
\label{eq.cond}
\end{align}
where $N$ is the total particle number, and $p(\bm{r})$ is the local pressure in a trap. Although our uniform calculation cannot completely describe this trapped geometry, noting that the left hand side of Eq. (\ref{eq.cond}) represents three times the grand potential per particle, we plot in Fig. \ref{fig10}(b) the calculated shear viscosity in SCTMA, imposing the condition $3P/N = \ef$ at each interaction strength (where $P$ is the pressure). This figure shows that our result semi-quantitatively explain the observed $\eta(T)$, which takes the minimum at $(k_{\rm F}a_s)^{-1}\sim 0.25$. This makes us expect that the minimum of the ratio $\eta/s$ may also exist away from the unitarity limit, although we need to calculate the entropy density $s$ to confirm this, which remains as our future problem.
\par
\section{Summary}
\par
To summarize, we have discussed the shear viscosity $\eta(T)$ in the BCS-BEC crossover regime of an ultracold Fermi gas. Including strong-coupling effects within the framework of the self-consistent $T$-matrix approximation (SCTMA), we have evaluated this transport coefficient in the normal state above the superfluid phase transition temperature $T_{\rm c}$.
\par
In the weak-coupling BCS regime ($(k_{\rm F}a_s)^{-1}\lesssim -0.5$), we found that $\eta(T)$ exhibits a dip structure, which is followed by a peak, as one decreases the temperature from $T_{\rm F}$. As the background physics of this, we pointed out that, while the former dip can be understood as the crossover from the classical regime to the Fermi degenerate regime, the latter peak is associated with the enhancement of pairing fluctuations near $T_{\rm c}$. Indeed, the latter peak temperature $T_{\rm peak}$ is closed to the pseudogap temperature $T^*$, which is defined as the temperature below which strong pairing fluctuations bring about a BCS-state like dip structure around the Fermi level in the single-particle density of states, in spite of the vanishing superfluid order parameter above $T_{\rm c}$.
\par
This non-monotonic temperature dependence once disappears in the unitary regime ($-0.5\lesssim (k_{\rm F}a_s)^{-1}\lesssim 0.7$), where $\eta(T)$ monotonically decreases with decreasing the temperature down to $T_{\rm c}$. This behavior was shown to be consistent with the recent experiment on a $^6$Li unitary Fermi gas.
\par
As one enters the strong-coupling BEC regime ($(k_{\rm F}a_s)^{-1}\gesim 0.7$),  $\eta(T)$ again exhibits a monotonic dip structure as a function of the temperature. In this case, the dip temperature $T_{\rm dip}^{\rm BEC}$ is close to the binding energy $E_{\rm bind}$ of a two-body bound molecules. This indicates that the rapid increase of $\eta(T)$ below $T_{\rm dip}^{\rm BEC}$ is due to the appearance of long-lived molecular bosons overwhelming thermal dissociations.
\par
The shear viscosity diverges in both the BCS and BEC limits, so that $\eta$ is expected to become minimum in the unitary regime. We have numerically confirmed this expectation; however, the minimum of $\eta(T)$ is obtained, not at the unitarity, but slightly in the BEC side. This deviation from the unitarity limit is consistent with the observed interaction dependence of this quantity in a $^6$Li Fermi gas.
\par
We finally note that the present work still has room for improvement. First, we could not calculate $\eta$ in the BEC regime near $T_{\rm c}$ because the numerical analytic continuation by the Pad\'e approximation did not work well there. To avoid this difficulty, an alternative approach which does not need the analytic continuation might be useful. Regarding this, the real-time formalism based on the Keldysh Green's function is promising. Second, SCTMA only includes effect of the molecular interaction in the BEC regime within the mean-field level, so that the molecular lifetime coming from inter-molecular scatterings is ignored. To improve this, at least, the second-order self-energy diagram in terms of this effective interaction must be taken into account. It is an interesting future challenge to explore how to incorporate such molecule-molecule scatterings into the present approach. Since we have only treated the normal state, the extension to the superfluid phase also remains as our future challenge. Since the shear viscosity has recently attracted much attention in various fields in connection to the KSS conjecture, our results would contribute to the further understanding of strongly interacting fermions, from the viewpoint of transport properties.
\par
\par
\acknowledgments
\par
We thank H.Tajima and R.Hanai for discussions. This work was supported by KiPAS project in Keio University. D.K. was supported by KLL Ph. D. Program Research Grant from Keio University. Y.O. was supported by a Grant-in-aid for Scientific Research from MEXT and JSPS in Japan (No.JP18K11345, No.JP18H05406, and No.JP19K03689).
\par
\par
\appendix
\par
\section{Expression for $\hat{\Pi}_{x,y}$ in the BCS model}
\label{app:derivation_Pi}
\par
The stress tensor operator $\hat{\Pi}_{i,j}$ ($i,j=x,y,z$) is defined from the conservation law in terms of the momentum flux,
\begin{align}
\partial_t {\hat J}_i + \sum_j \partial_j {\hat \Pi}_{i,j} = 0,
\label{eq_momentum_consv}
\end{align}
where ${\hat {\bm J}} = \sum_{\sigma}[\psi_{\sigma}^{\dag} (\nabla \psi_{\sigma})-(\nabla \psi_{\sigma}^{\dag}) \psi_{\sigma}]/(2i)$ is the mass current operator, with $\psi_{\sigma}$ being the Fermi field operator. For the ordinary Hamiltonian consisting of the kinetic term and a short-range interaction potential $V(r_{12})\equiv V(|{\bf r}_1-{\bf r}_2|)$, one has\cite{Martin:1959aa},
\begin{align}
\hat{\Pi}_{i,j}(\bm{r}) 
&= \sum_{\sigma} \frac{1}{2m} \left[ (\nabla_i \psi^{\dag}_{\sigma})(\nabla_j \psi_{\sigma}) + (i\leftrightarrow j)
-\frac{1}{2} \nabla_i \nabla_j (\psi^{\dag}_{\sigma} \psi_{\sigma})
\right]
\nonumber
\\
&-
\int d\bm{r}_{12} \frac{(\bm{r}_{12})_i (\bm{r}_{12})_j}{r_{12}^2}
\psi^{\dag}_{\up}\left(\bm{r}+\frac{1}{2} \bm{r}_{12}\right) \psi^{\dag}_{\dwn}\left(\bm{r}-\frac{1}{2} \bm{r}_{12}\right)
\nonumber
\\
&\times
\left[ r_{12} \frac{\partial V(r_{12})}{\partial r_{12}} \right]
\psi_{\dwn}\left(\bm{r}-\frac{1}{2} \bm{r}_{12}\right) \psi_{\up}\left(\bm{r}+\frac{1}{2} \bm{r}_{12}\right).
\label{eq_stress_tensor_def}
\end{align}
To obtain the detailed expression for $r_{12}(\partial V(r_{12})/\partial r_{12})$ in the present BCS model in Eq. (\ref{eq.1.0}), we evaluate the pressure $P$ by using the Hellmann-Feynman theorem,
\begin{align}
P = - \left\langle \frac{\partial H}{\partial V} \right\rangle.
\label{eq.a1}
\end{align}
We assume that a gas is confined in a box with the size $V=L\times L\times L$, and the periodic boundary condition is imposed. In this case, the momentum ${\bm p}$ is quantized as ${\bm p}=2\pi{\bm n}/L$ (where $\bm{n}=(n_x,n_y,n_z)$ is the quantum number), which depends on the system size $L$. In addition, the cutoff momentum $p_{\rm c}=2\pi|n_{\rm c}|/L$ also depends on $L$ (where $n_{\rm c}$ is a large quantum number). Keeping these in mind, varying the system size as $L\to L(1+\epsilon)$, we obtain the pressure from Eq. (\ref{eq.a1}) as,
\begin{align}
PV &
= - \left\langle \lim_{\epsilon \to 0} \frac{H(L(1+\epsilon)) - H(L)}{(1+\epsilon)^3 - 1} \right\rangle
\nonumber
\\
&=
\frac{2}{3} \sum_{\bp,\sigma} \varepsilon_{\bp} \left\langle c^{\dag}_{\bp,\sigma} c_{\bp,\sigma} \right\rangle
\nonumber
\\
&-
\left[
U - U^2 \frac{m p_{\rm c}}{6\pi^2}
\right]
\frac{1}{V} \sum_{\bp,\bp',\bq} \left\langle c^{\dag}_{\bm{p}+\bm{q},\uparrow} c^{\dag}_{\bm{p}'-\bm{q},\downarrow} c_{\bm{p}',\downarrow} c_{\bm{p},\uparrow} \right\rangle.
\label{eq.a.3}
\end{align}
We briefly note that, when we use Eq. (\ref{eq_renormalization}) to remove the cutoff $p_{\rm c}$, Eq. (\ref{eq.a.3}) is found to reproduce the Tan's pressure relation \cite{Tan:2008aa,Tan:2008ab,Tan:2008ac}, $PV = 2E/3+C/(12\pi m a_s)$, where $E$ is the internal energy, and $C = (m^2 U^2/V)\sum_{\bp,\bp',\bq} \left\langle c^{\dag}_{\bm{p}+\bm{q},\uparrow} c^{\dag}_{\bm{p}'-\bm{q},\downarrow} c_{\bm{p}',\downarrow} c_{\bm{p},\uparrow} \right\rangle$ is the Tan's contact\cite{Braaten:2008aa}.
\par
Comparing Eq. (\ref{eq.a.3}) with the relation between the pressure and the stress tensor,
\begin{align}
PV=\frac{1}{3} \int d\bm{r} \sum_{i} \langle \hat{\Pi}_{i,i}(\bm{r}) \rangle,
\end{align}
one finds, 
\begin{align}
\frac{(\bm{r}_{12})_i (\bm{r}_{12})_j}{r_{12}^2} \left[ r_{12} \frac{\partial V(r_{12})}{\partial r_{12}} \right]
= \delta_{i,j} \left[ \frac{2}{3} U - U^2\frac{m}{12\pi \as} \right] \delta(\bm{r}_{12}).
\end{align}
Substituting this into Eq. (\ref{eq_stress_tensor_def}), we reach
\begin{align}
\hat{\Pi}_{i,j}(\bm{r}) 
&= \sum_{\sigma} \frac{1}{2m}
\left[ (\nabla_i \psi^{\dag}_{\sigma})(\nabla_j \psi_{\sigma}) 
+ (i\leftrightarrow j)
-\frac{1}{2} \nabla_i \nabla_j (\psi^{\dag}_{\sigma} \psi_{\sigma})
\right]
\nonumber
\\
&-\delta_{i,j} \left[ \frac{2}{3} U - U^2\frac{m}{12\pi \as} \right] \psi^{\dag}_{\up}\left(\bm{r}\right) \psi^{\dag}_{\dwn}\left(\bm{r}\right)
\psi_{\dwn}\left(\bm{r}\right) \psi_{\up}\left(\bm{r}\right).
\label{eq_stress_tensor}
\end{align}
\par
We note that, although $\hat{\Pi}_{i,j}' \equiv \hat{\Pi}_{i,j} + \delta\hat{\Pi}_{i,j}$ with any symmetric tensor $\delta\hat{\Pi}_{i,j}$ satisfying $\sum_{j} \partial_j \delta\hat{\Pi}_{i,j} = 0$ also satisfies Eq. (\ref{eq_momentum_consv}), this non-uniqueness doesn't affect the shear viscosity\cite{Taylor:2010aa}. We also note that the interaction term in Eq. (\ref{eq_stress_tensor}) is for the case when Eq. (\ref{eq_renormalization}) is used. When one employs another relation between $-U$ and a renormalized quantity, e.g., the dimensional regularization, the interaction term in Eq. (\ref{eq_stress_tensor}) is replaced by the corresponding form. The expression for th shear stress tensor in the case of the dimensional regularization is given in Ref. \cite{Fujii:2018aa}.
\par
\par
\section{Shear viscosity obtained from the Boltzmann equation}
\label{app:Boltzmann}
\par
This appendix summarizes how to calculate $\eta$ in the Boltzmann-equation approach. For a two-component Fermi gas, the Boltzmann equation is given by
\begin{align}
\frac{\partial f(\bm{r},\bm{p},t)}{\partial t} + \frac{\bm{p}}{m} \cdot \frac{\partial f(\bm{r},\bm{p},t)}{\partial \bm{r}} = -C[f]
\label{eq.b0}
\end{align}
where $f(\bm{r},\bm{p},t)$ is a quasi-classical distribution function, and
\begin{align}
C[f] &= \sum_{\bp_1,\bp',\bp'_1} \frac{d \sigma}{d \Omega}
(2\pi) \delta(\varepsilon_{\bp} + \varepsilon_{\bp_1} - \varepsilon_{\bp'} - \varepsilon_{\bp'_1})
\delta_{\bp+\bp_1,\bp'+\bp'_1}
\nonumber
\\
&\times
\left[
f f_1 (1 - f') (1 - f'_1) - (1 - f) (1 - f_1) f' f'_1
\right]
\label{eq.b1}
\end{align}
is the collision integral, with $d \sigma / d \Omega$ being the differential cross section, $f_1=f(\bm{r},\bp_1,t)$, $f'=f(\bm{r},\bp',t)$, and $f'_1=f(\bm{r},\bp'_1,t)$. 
\par
To evaluate the shear viscosity $\eta$, we consider the current-flowing steady state, where the fluid velocity $\bm{u}(\bm{r})=(u_x(y),0,0)$, $u_x$, as well as the gradient $\partial u_x / \partial y$, are assumed to be small. In this case, the shear viscosity $\eta$ is related to the shear stress $\Pi_{x,y}$ as,
\begin{align}
\Pi_{x,y} = - \eta \frac{\partial u_x}{\partial y}.
\label{eq.b3}
\end{align}
$\Pi_{x,y}$ is also related to the quasi-classical distribution function as,
\begin{align}
\Pi_{x,y}(\bm{r},t) = 2 \sum_{\bp} \frac{p_xp_y}{m}f(\bm{r},\bm{p},t),
\label{eq.b2}
\end{align}
\par
To calculate Eq. (\ref{eq.b2}), we divide the distribution function $f=f_{\rm loc}+\delta f$ into the sum of the local equilibrium distribution function,
\begin{equation}
f_{\rm loc}(\bm{r},\bm{p}) = \frac{1}{e^{(\xi_{\bm{p}} - \bm{u}(\bm{r}) \cdot \bm{p})/T} +1},
\end{equation}
and the deviation $\delta f$ from it. Following the standard technique, we parametrize the latter as
\begin{align}
\delta f(\bm{r},\bm{p}) = 
-\frac{1}{T}f_{\rm loc}(\bm{r},\bm{p}) 
\left[
1 - f_{\rm loc}(\bm{r},\bm{p})
\right]
\left( \frac{\partial u_x}{\partial y}\right) 
\Phi(\bm{r},\bm{p}).
\end{align}
Substituting $f=f_{\rm loc}+\delta f$ into Eq. (\ref{eq.b2}), and using Eq. (\ref{eq.b3}), we obtain\cite{Massignan:2005aa,Bruun:2005aa,Bruun:2007aa,Enss:2012ab,Bruun:2012aa,Bluhm:2014aa}
\begin{align}
\eta =  - 2 \sum_{\bp} \frac{p_xp_y}{m} \frac{\partial f_{\rm F}(\xi_{\bp})}{\partial \xi_{\bp}} \Phi(\bm{p})
\label{eq_eta_boltzmann}
\end{align} 
\par
The function $\Phi(\bm{p})$ is obtained from the linearized Boltzmann equation, which is derived by substituting $f=f_{\rm loc}+\delta f$ into Eq. (\ref{eq.b0}) and Eq. (\ref{eq.b1}), and retaining terms with $O(\partial u_x / \partial y)$. The result is
\begin{align}
\nonumber
\frac{p_xp_y}{m} f^0[1-f^0]
&=
\sum_{\bp_1,\bp',\bp'_1} \frac{d \sigma}{d \Omega}
(2\pi) 
\delta(\varepsilon_{\bp} + \varepsilon_{\bp_1} - \varepsilon_{\bp'} - 
\varepsilon_{\bp'_1})
\nonumber
\\
&\times\delta_{\bp+\bp_1,\bp'+\bp'_1} 
f^0 f^0_1 [1 - {f^0}'][1-{f_1^0}'] 
\nonumber
\\
&\times\left[\Phi(\bp) + \Phi(\bp_1) - \Phi(\bp') - \Phi(\bp'_1) \right].
\label{eq_linearized_boltzmann}
\end{align}
Here, $f^0=f_{\rm F}(\xi_{\bp})$, $f_1^0=f_{\rm F}(\xi_{\bp_1})$, ${f^0}'=f_{\rm F}(\xi_{\bp'})$, ${f^0_1}'=f_{\rm F}(\xi_{\bp'_1})$.
\par
Following the previous works\cite{Massignan:2005aa,Bruun:2005aa,Bruun:2007aa,Enss:2012ab,Bruun:2012aa,Bluhm:2014aa}, we set $\Phi(\bp) = A p_x p_y / m$, where $A$ is a constant. Substituting this into the linearized Boltzmann equation (\ref{eq_linearized_boltzmann}), and multiplying $\sum_{\bp} p_xp_y/m$ in both the sides of this equation, we have
\begin{align}
\nonumber
\sum_{\bp} \left(\frac{p_xp_y}{m}\right)^2 f^0[1-f^0]
&=A
\sum_{\bp,\bp_1,\bp',\bp'_1} \frac{d \sigma}{d \Omega} 
\delta_{\bp+\bp_1,\bp'+\bp'_1} 
\nonumber
\\
&\hskip-25mm\times
(2\pi) \delta(\varepsilon_{\bp} + \varepsilon_{\bp_1} - \varepsilon_{\bp'} - \varepsilon_{\bp'_1})
f^0 f_1^0 [ 1 - {f^0}' ][ 1 - {f_1^0}'] 
\nonumber
\\
&\hskip-25mm\times\frac{p_xp_y}{m} \left[ \frac{p_xp_y}{m} + \frac{p_{1,x}p_{1,y}}{m} - \frac{p'_xp'_y}{m} - \frac{p'_{1,x}p'_{1,y}}{m} \right],
\end{align}
Once $A$ is determined, $\eta$ is immediately obtained as
\begin{align}
\eta =  - 2 \sum_{\bp} \left( \frac{p_xp_y}{m} \right)^2 
\frac{\partial f_{\rm F}}{\partial \xi_{\bp}}A.
\label{eq.b5}
\end{align}
In Fig. \ref{fig5}, `BE' shows the result obtained from Eq. (\ref{eq.b5}), where the SCTMA chemical potential $\mu(T)$ in Fig. \ref{fig2} is used.
\par
In $^{40}$K and $^6$Li Fermi gases, the effective range of the Feshbach-induced tunable interaction is negligibly small. In this case, the differential cross section has the resonance form\cite{Massignan:2005aa,Bruun:2005aa},
\begin{align}
\frac{d\sigma}{d\Omega} = \frac{\as^2}{1+(k\as)^2},
\end{align}
where $k = |\bp - \bp_1|/2$. In the classical regime ($T\gg T_{\rm F}$), one obtains
\begin{align}
\eta = \frac{5\sqrt{\pi m T}}{8\bar{\sigma}},
\label{eq.b6}
\end{align}
where
\begin{align}
\bar{\sigma} = \frac{4\pi\as^2}{3} \int_0^{\infty} dx x^7 \frac{e^{-x^2}}{1+x^2 T m\as^2}.
\label{eq.b7}
\end{align}
Equation (\ref{eq.b6}) is reduced to\cite{Massignan:2005aa,Bruun:2005aa}
\begin{align}
&\eta = \frac{15}{32\sqrt{\pi}}(mT)^{\frac{3}{2}}
~~~~~~(T\gg 1/(ma_s^2)),
\\
&\eta = \frac{5}{32\sqrt{\pi}\as^2}(mT)^{\frac{1}{2}}
~~~( T\ll 1/(ma_s^2)). 
\end{align}
\par
In the Fermi degenerate regime ($T\ll T_{\rm F}$), in the weak coupling BCS regime, Eq. (\ref{eq.b5}) gives,\cite{Massignan:2005aa,Bruun:2005aa}
\begin{equation}
\eta = \frac{3n}{8\pi (\kf \as)^2}
\left(\frac{\Tf}{T}\right)^2.
\end{equation}
\par
We briefly explain the background physics of these results in the classical and quantum regime. In the simple expression $\eta \sim n l_{\rm mfp} \bar{p}$\cite{Pitaevskii:1981aa}, the mean free path $l_{\rm mfp} \sim 1/(n\sigma)$\cite{Pitaevskii:1981aa} is related to the collisional cross section $\sigma$ (where $n$ is the number density), and the averaged momentum ${\bar p}$ is estimated as ${\bar p}\sim \sqrt{mT}$. When the thermal de-Broglie wave length $\lambda_T = 1/\sqrt{mT}$ is much larger than the scattering length $|\as|$, the area where collisions occur is estimated as $\sim\as^2$, giving $\sigma\propto \as^2$. In this case, one obtains $\eta\propto T^{1/2}$. When $\lambda_T\ll |a_s|$, on the other hand, the area where collisions occur is estimated as $\sim \lambda_{T}^2$, because collision cannot occur when wave-functions of particles don't overlap. In this case, one has $\sigma\propto 1/(mT)$, leading to $\eta\propto T^{1.5}$.
\par
When we still use the expression $\eta \sim n l_{\rm mfp} \bar{p}$ in the Fermi degenerate regime, one may take ${\bar p}\sim k_{\rm F}$, and $l_{\rm mfp} \sim v_{\rm F}\tau\propto T^{-2}$, where $v_{\rm F}=k_{\rm F}/m$ is the Fermi velocity, and $\tau\propto T^2$ is the lifetime of quasi-particles originating from a particle-particle interaction. These give $\eta(T)\propto T^{-2}$, as expected.
\par
\par
\section{Derivation of Eq. (\ref{eq.eta_BEC})}
\label{app:derivation_eq.eta_BEC}
\par
The bare molecular three-point vertex function $S_{0;x,y}$ in Fig. \ref{fig8}(c) consists of three Fermi Green's functions as
\begin{align}
S_{0;x,y}(\bm{q},i\nu_m,i\nu_m + i\nu_l)&= 
T\sum_{\bm{p},\omega_n}  \frac{p_x p_y}{m} G(\bm{q}-\bm{p},i\nu_m - i\omega_n)
\nonumber
\\
&\times
G(\bm{p},i\omega_n)G(\bm{p},i\omega_n+i\nu_l).
\label{eqC1}
\end{align}
We evaluate Eq. (\ref{eqC1}) in the BEC regime within the replacement of the SCTMA dressed Green's function $G$ by the bare one $G_0$. The result is, at $\nu_m=\nu_l=0$,
\begin{align}
S_{0;x,y}(\bm{q},0,0)
\simeq
-\frac{m^2 \as}{8 \pi} \frac{q_x q_y}{4m},
\label{eq.c2}
\end{align}
Approximately using Eq. (\ref{eq.c2}) for $S_{0;x,y}$ in the shear-stress response function $\Xi(i\nu_l)$ in Fig. \ref{fig8}(e)\cite{note2}, we obtain
\begin{equation}
\Xi(i\nu_l) = - T\sum_{\bm{q},\nu_m} 
\left(
\frac{q_x q_y}{2m}
\right)^2
G_{\rm B}(\bm{q},i\nu_m) G_{\rm B}(\bm{q},i\nu_m + i\nu_l),
\label{eq.C3}
\end{equation}
where the molecular Bose Green's function has the form,
\begin{equation}
G_{\rm B}({\bm q},i\nu_m)=
\frac{1}{i\nu_m-\xi_{\bm q}^{\rm B}+i \tau_{\rm B}^{-1}{\rm sgn}(\nu_m)}.
\label{eq.C4}
\end{equation}
Carrying out the $\nu_m$-summation, we execute the analytic continuation $i\nu_l\to\omega_+$ in Eq. (\ref{eq.C3}). Then, we obtain from Eq. (\ref{eq.100}),
\begin{align}
\eta = 
-\sum_{\bm{q}} 
\left(
\frac{q_x q_y}{2m}
\right)^2
\int \frac{dz}{\pi}
\frac{\partial n_{\mathrm{B}}(z)}{\partial z}
\left[
{\rm Im}[G_{\rm B}({\bm q},i\nu_m\to z+i\delta]
\right]^2.
\label{eq.C5}
\end{align}
Using the fact that $\tau_{\rm B} = \lambda e^{E_{\rm bind}/(2T)}$ is very large because of the exponential factor, we further employ the approximation,
\begin{align}
\left[
{\rm Im}[G_{\rm B}({\bm q},i\nu_m\to z+i\delta]
\right]^2
=\frac{\pi}{2} \tau_{\rm B} \delta(z - \xi_{\bm{q}}^{\mathrm{B}}).
\end{align}
Substitution of this into Eq. (\ref{eq.C5}) gives Eq. (\ref{eq.eta_BEC}).
\par
\par


\begin{thebibliography}{9}
\bibitem{note1} In Fig. \ref{fig5}(a), we see difference between the SCTMA and Boltzmann-equation results at high temperatures when $(k_{\rm F}a_s)^{-1}=-1$, although effects of pairing fluctuations are considered to be weak there. We also note that the same problem also occurs for stronger interactions, although we do not show the result here. At this stage, although the reason for this discrepancy is unclear, it is considered that our numerical analytic continuation explained in Sec. \ref{sec:formulation} may underestimate $\eta$ in such high temperature region ($T\gesim T_{\rm F}$).
\bibitem{note2} When the inter-molecular interaction $U_{\mathrm{B}}$ is approximated by the $s$-wave one as Eq. (\ref{eq.intB}), the second term of the right hand side in Fig. \ref{fig8}(e) vanishes because the three-point vertex $S_{x,y}$ is proportional to $q_xq_y$.
Thus, in this case, the three-point vertex $S_{x,y}$ is reduced to the lowest-order one $S_{0;x,y}$.
\end{thebibliography}
\end{document}